\documentclass[conference]{IEEEtran}
\IEEEoverridecommandlockouts
\usepackage{cite}
\usepackage{amsmath,amssymb,amsfonts}
\usepackage{algorithmic}
\usepackage{graphicx}
\usepackage{textcomp}
\usepackage{hyperref}
\usepackage{xcolor}
\def\BibTeX{{\rm B\kern-.05em{\sc i\kern-.025em b}\kern-.08em
    T\kern-.1667em\lower.7ex\hbox{E}\kern-.125emX}}

\usepackage{array}
\usepackage{mdwmath}
\usepackage{mdwtab}
\usepackage{eqparbox}
\usepackage{url}
\usepackage{supertabular}                                             
\usepackage{caption}
\usepackage{subcaption}
\usepackage{float}
\usepackage{booktabs}
\usepackage{multirow}
\usepackage{graphicx}
\usepackage{color}
\usepackage{tabularray}
\usepackage{hhline}
\usepackage{wasysym}
\usepackage{rotating}
\graphicspath{ {Figures/} }
\UseTblrLibrary{diagbox}

\begin{document}

\title{Lack of Systematic Approach to Security of IoT Context Sharing Platforms\\
%\thanks{Identify applicable funding agency here. If none, delete %this.}
}

\author{
    \IEEEauthorblockN{Mohammad Goudarzi\IEEEauthorrefmark{1}\IEEEauthorrefmark{3}, Arash Shaghaghi\IEEEauthorrefmark{1}\IEEEauthorrefmark{3}, Simon Finn\IEEEauthorrefmark{2}, Sanjay Jha\IEEEauthorrefmark{1}\IEEEauthorrefmark{3}}
    \IEEEauthorblockA{\IEEEauthorrefmark{1}School of Computer Science and Engineering, UNSW Sydney, Australia}
    \IEEEauthorblockA{\IEEEauthorrefmark{3}Cyber Security Cooperative Research Centre}
    \IEEEauthorblockA{\IEEEauthorrefmark{2}Cisco \\
    }

}

%Contact: moh.goudarzi90@gmail.com
\maketitle

\begin{abstract}
IoT context-sharing platforms are an essential component of today's interconnected IoT deployments with their security affecting the entire deployment and the critical infrastructure adopting IoT. We report on a lack of systematic approach to the security of IoT context-sharing platforms and propose the need for a methodological and systematic alternative to evaluate the existing solutions and develop `secure-by-design' solutions. We have identified the key components of a generic IoT context-sharing platform and propose using MITRE ATT\&CK for threat modelling of such platforms.
\end{abstract}


\section{Introduction}
Context-aware systems are implemented in Internet of Things (IoT) environments to detect the environment and to deliver an appropriate response to both the user and the application \cite{de2020context}. The information that is generated by the IoT devices is analysed by these systems, which then provide a high-level semantic and transform it into context information. When attempting to define the condition of an environment, knowledge about the context is utilised. The context information of an environment can be produced by different elements, such as a user, an application, or a device. In IoT deployments, context information is typically stored locally and not shared, which restricts the return on investment and value of these deployments \cite{ramachandran2019towards}. 
\par 
Context sharing among IoT deployments is a fundamental functioning requirement since it enables computers to comprehend context information across a variety of environments and applications. Let us examine the scenario of smart transportation, where IoT devices and the data they generate are utilised in various application areas such as smart traffic and smart station. Each of these applications necessitates distinct systems, and the sharing of context allows these systems to comprehend contextual information across diverse settings that may vary in terms of format, data kinds, and specifications. Context-sharing allows systems in pervasive computational contexts to share information and work together effectively. The importance of context-sharing platforms has motivated active research in academia and industry. Authors in \cite{de2020context} provide a summary of research conducted in academic literature. Prominent industry and government initiatives, such as the EU Horizon 2020 and EU FP7, have particularly undertaken substantial efforts to develop and construct functional frameworks for sharing contextual information in IoT deployments. We will briefly review them in Section \ref{sec:IoTContext-sharingIndustryProjects} of this paper.
\par
The security of context information exchanges possesses distinct properties and necessitates certain criteria. For example, if a malicious attacker acquires unencrypted information via a communication channel, comprehending its significance may prove challenging without the appropriate contextual information. Nevertheless, contextual information frequently conveys a particular occurrence in a semantic manner, so enhancing its comprehensibility to a potential attacker. Most context-sharing platforms lack a comprehensive and integrated examination of security vulnerabilities \cite{de2020context}. This gap is substantial and presents a serious barrier to the durability and dependability of large-scale IoT deployments. Noting that the security of critical infrastructure adopting IoT can be targeted by adversaries through the context-sharing platforms enabling their ecosystem. The interconnected and complex deployment models of context-sharing platforms challenges securing them against both traditional (e.g., DoS, information leakage) and more recent attacks targeting critical infrastructure such as stealthy infiltration attacks \cite{rieger2023argus} and energy consumption attacks \cite{hlavacs2011energy}. In this paper, we raise the importance of taking a systematic approach to the security of IoT context-sharing platforms. We first report on our review of selected well-known industry projects in this domain including EU FP7 and HORIZON 2020. Based on these projects, we identify the main elements of a typical context-sharing platform and the phases involved. We then suggest using MITRE Adversarial Tactics, Techniques, and Common Knowledge (ATT\&CK) \cite{Mitre}, which is a well-known, actively developed, and industry-supported threat modelling framework, to systematically analyse IoT context-sharing platforms. To prove the feasibility of conducting such an analysis, we report on one specific tactic across all components and phases. We conclude by outlining the importance of conducting a systematic and methodological threat modelling for a context-sharing platform to analyse proposed solutions and help design the next generation of these platforms to ensure they are `secure by design'.

\section{Background}
This section describes the background information about context-sharing platforms and MITRE ATT\&CK.
\subsection{Context Sharing Platform}
In order to give context information, a system must adhere to a series of sequential actions. Context-sharing platforms usually have four stages, including context extraction, context modelling, context reasoning, and context distribution \cite{zhang2021middleware}. Context extraction involves collecting raw data from a sensor, database, or the environment. The context modelling stage transforms the data into a specific format (e.g., key-value pairs, ontology) to prepare it for the reasoning stage. The process of reasoning is the initial and most important step in the life-cycle of a context. It converts the information into a context, making it comprehensible to the end users via different strategies such as ontology and data fusion. Context distribution is a simple and direct process, where its primary function is to disseminate contextual information through either direct queries or subscriptions.
\subsection{MITRE ATT\&CK}
MITRE is a globally accessible knowledge repository that captures insights into adversarial tactics and methodologies based on real-world observations. MITRE ATT\&CK serves as a structured model and knowledge base for understanding adversary behaviour, encompassing various stages of the attack life-cycle and the generic platforms they use to target systems. It has two main elements: tactics representing an attacker’s objectives and techniques displaying how these objectives can be achieved. MITRE ATT\&CK is categorized into three primary domains, including enterprise, mobile, and industrial control systems. Besides, the ATT\&CK matrix provides mitigations that are security concepts and sets of tools designed to prevent the effective execution of specific techniques or sub-techniques by adversaries.
\section{IoT Context-sharing Industry Projects}
\label{sec:IoTContext-sharingIndustryProjects}
We have identified two prominent industrial initiatives focused on IoT context-sharing platforms, namely HORIZON 2020 and EU FP7. Table 1 presents the list of these projects. We have thoroughly examined the deliverables that are publically accessible for both projects.
Three EU FP7 projects mostly concentrate on context-sharing platforms. The primary objective of the Internet of Things Architecture (IoT-A) \cite{europaInternetThings} is to establish a comprehensive architectural reference model that ensures interoperability inside the IoT system. It aims to provide design principles and recommendations for protocols, interfaces, and algorithms. The goal is to provide a data format that is suitable for situations with limited resources, in order to reduce network traffic and the frequency of interactions, therefore improving interoperability in a more effective manner. The objective of the COllaboration and INteroperability for networked enterprises (COIN) \cite{europaCOllaborationINteroperability} project is to facilitate the integration of both existing and new enterprise interoperability and collaboration services. COIN provides a web solution that uses semantics to ensure compatibility and cooperation across different systems. It uses an approach based on ontologies to process and understand the meaning of data. Furthermore, COIN incorporates a knowledge-based system that stores information pertaining to diverse elements, such as devices and resources. The primary objective of the Interoperability of Data and Procedures The main objective of the large-scale multinational disaster Response Actions (IDIRA) \cite{europaInteroperabilityData} is to enhance the sharing of information and interoperability of services, as well as provide decision assistance to local and international entities involved in crisis management.
\begin{table}[t]
\centering
\caption{IoT Context-sharing Platform (Industry Projects)}
\label{tab:evaluation:securityconsideration-Industry}
\resizebox{\linewidth}{!}{%
\begin{tblr}{
  cells = {c},
  cell{2}{1} = {r=3}{},
  cell{5}{1} = {r=6}{},
  vlines,
  hline{1-2,5,11-13} = {-}{},
  hline{3-4,6-10} = {2-6}{},
}
Organization    & Project    & Main Domain            \\
EU FP7          & IOT-A      & IoT                    \\
                & COIN       & Industry                \\
                & IDIRIA     & Crisis Management       \\
{HORIZON\\2020} & Inter-IOT  & IoT                      \\
                & FIESTA-IoT & {IoT - Environment\\~as a Service (EaaS)} \\
                & Wise-IoT   & IoT                      \\
                & SEMIoTICS  & IoT                       \\
                & BIG IoT    & IoT                       \\
                & SymbIoTe   & IoT                       
\end{tblr}
}
\end{table}
\par 
Out of the HORIZON 2020 projects, six specifically focus on context-sharing platforms. The primary objective of the compatibility of Heterogeneous IoT systems (Inter-IoT) \cite{europaInteroperabilityHeterogeneous} project is to guarantee compatibility between different IoT systems that have varying characteristics. It works together with other projects to ensure compatibility across different layers, such as device, networking, middleware, application service, data and semantics, integrated IoT platform, and business levels. INTER-IoT has developed the Generic Ontology for IoT Platforms to simplify semantic matching in IoT situations and make the sharing of context more efficient at the data and semantics level. Inter-IoT implemented its framework in two specific areas of application, namely e-health and port transportation. The primary objective of the Federated Interoperable Semantic IoT/cloud Testbeds and Applications (FIESTA) \cite{europaFederatedInteroperable} is to enhance IoT experimentation by enabling the connectivity and compatibility of different IoT systems, platforms, and testbeds. The system utilises a common ontology to maintain semantic coherence across different providers. It also leverages a standardised API to enable communication and provide access to information for IoT systems connected to it. The objective of the Worldwide Interoperability for SEmantics IoT (Wise-IoT) \cite{europaWorldwideInteroperability} project is to provide interoperability in the exchange of context information. The Global IoT Services layer is introduced, which is characterised by semantic interoperability to ensure reliability and end-to-end security. The Morphing Mediation Gateway component plays a crucial role in translating different protocols and data formats, while also accommodating varied ontologies. The objective of the Smart End-to-end Massive IoT Interoperability, Connectivity and Security (SEMIoTICS) \cite{europaSmartEndtoend} project is to offer a pattern-driven approach to achieve semantic interoperability in IoT environments. This approach enables the seamless integration and adaptation of various smart objects across different platforms. SEMIoTICS tackles the important issue of guaranteeing secure and dependable actuation in IoT and industrial IoT applications by creating a pattern-driven framework. It encodes pre-existing relationships between security, privacy, dependability, and interoperability, making it easier to smoothly integrate smart objects. The Bridging the Interoperability Gap of the Internet of Things (Big IoT) \cite{europaBridgingInteroperability} offers an Application Programming Interface (API), which is a web-based interface specifically developed to improve interoperability between IoT systems. Additionally, it encompasses the creation of the BIG IoT Marketplace, which facilitates the efficient sharing and monetization of data among platforms. This marketplace enables service, application providers, and platform operators to generate revenue from their assets, thereby lowering the obstacles for developers and promoting a more inclusive ecosystem in the IoT field. SymbIoTe \cite{europaSymbiosisSmart} is an IoT orchestration middleware that offers a cohesive view across several IoT platforms. This technology allows for safe and protected entry to both real and virtualized IoT resources. It also simplifies the process of finding and managing these resources in a hierarchical manner across various platforms. Furthermore, it promotes the formation of a network of IoT controllers and resources, which in turn encourages cooperative sensing and actuation tasks. The orchestration middleware leverages established protocols and interfaces, integrating proprietary platforms from industrial partners and open-source platforms such as OpenIoT to guarantee interoperability and compatibility across IoT contexts.
\section{Threat Modelling of IoT Context Sharing Platforms}
In this section, we will outline the primary elements of the context-sharing platforms and explain how MITRE ATT\&CK can be employed for threat modelling of the IoT context-sharing platforms.
\subsection{Main Elements of Context-sharing Platforms}
In our analysis of real-world industry projects on IoT context-sharing platforms, we have identified the key elements that play an active role in these platforms. For this, we have abstracted the designs of various systems reported in various projects, including those reported in Table 1. These components include the Data Consumer, Hardware, Transmission, Operating System (OS), Application Software, Data, and Virtualization.
\par
Data Consumer refers to individuals who are either present within the defined environment or interact with the context-sharing platform. Hardware refers to all tangible devices and components within a context-sharing platform, except network devices. Examples encompass a variety of entities, including servers and sensors. Transmission encompasses the tangible components of a network, such as routers and switches, along with all the network protocols, including IP and routeing. OS encompasses any operating system, such as MacOS and Windows. Application refers to any software, service, or process that runs on an OS. Data encompasses all unprocessed information obtained from the surroundings or any contextual information that can be accessed within the context-sharing platform. Within this data category, we also incorporate the media used for data storage, such as a file. Virtualization refers to the creation of virtualized versions of previously mentioned elements, such as containers and virtual machines (VMs).
\subsection{Integrating ATT\&CK with a Context-sharing Platform}
%
%In order to comprehensively understand the security risks associated with IoT context-sharing platforms, it is crucial to conduct a systematic analysis of the main fundamental components, taking into account their specific qualities and vulnerabilities, and the main stages within these platforms.
%
\begin{table*}[]
\caption{Threat Modelling of IoT Context-sharing Platforms using MITRE ATT\&CK (Credential Access Tactic)}
\label{tab:credentialAccess}
\resizebox{\textwidth}{!}{%
\begin{tabular}{|ccccccc|}
\hline
\multicolumn{7}{|c|}{Credential Access} \\ \hline
\multicolumn{6}{|c|}{\begin{tabular}[c]{@{}c@{}}Context Extraction and\\ Contect Distribution\end{tabular}} & \begin{tabular}[c]{@{}c@{}}Context Modelling and\\ Context Distribution\end{tabular} \\ \hline
\multicolumn{1}{|c|}{Data Consumer} & \multicolumn{1}{c|}{Hardware} & \multicolumn{1}{c|}{Transmission} & \multicolumn{1}{c|}{OS} & \multicolumn{1}{c|}{Application} & \multicolumn{1}{c|}{Virtualization} & Data \\ \hline
\multicolumn{1}{|c|}{\begin{tabular}[c]{@{}c@{}}T1557(001,003)(E),\\ T1110(001-004)(E),\\ T1551(001,002,004,005)(E),\\ T1212(E),T1187(E),\\ T1556(001-003,005-008)(E),\\ T1621(E),T1040(E),\\ T1003(001-006,008)(E),\\ T1552(001,002,004,008)(E),\\ T1517(M),T1414(M),\\ T1634(001)(M),T1111(E),\\ T1417(001-002)(M),\\ T1056(001-004)(E)\end{tabular}} & \multicolumn{1}{c|}{\begin{tabular}[c]{@{}c@{}}T1110(001-004)(E),\\ T1056(001)(E),\\ T1111(E),T1040(E),\\ T1552(001,004)(E),\\ T1621(E),\\ T1551(004)(E)\end{tabular}} & \multicolumn{1}{c|}{\begin{tabular}[c]{@{}c@{}}T1557(002,003)(E),\\ T1556(004)(E),\\ T1110(001-004)(E),\\ T1111(E),\\ T1551(004)(E),\\ T1552(001,004)(E),\\ T1056(001,002)(E),\\ T1634(001)(M),\\ T1621(E),T1040(E),\end{tabular}} & \multicolumn{1}{c|}{\begin{tabular}[c]{@{}c@{}}T1110(001-004)(E),\\ T1551(001,002,004)(E),\\ T1056(001,002)(E),\\ T1003(001-006)(E),\\ T1558(001-004)(E),\\ T1552(001-004)(E),\\ T1417(001-002)(M)\\ T1621(E), T1212(E),\\ T1557(001)(E), T1649(E),\\ T1634(001)(M),T1040(E)\end{tabular}} & \multicolumn{1}{c|}{\begin{tabular}[c]{@{}c@{}}T1110(001-004)(E),T1212(E),\\ T1634(001)(M),T1111(E),\\ T1551(001-004)(E),T1621(E),\\ T1056(001-004)(E),T1539(E),\\ T1414(M), T1040(E),\\ T1417(001-002)(M),T1649(E),\\ T1003(004,007)(E),T1528(E),\\ T1517(M),T1635(001)(M),\\ T1552(001-004)(E)\end{tabular}} & \multicolumn{1}{c|}{\begin{tabular}[c]{@{}c@{}}T1110(001-004)(E),\\ T1551(001,002)(E),\\ T1606(001-002)(E),\\ T1056(001-003)(E),\\ T1040(E),T1528(E),\\ T1649(E),T1539(E),\\ T1212(E),T1621(E),\\ T1552(001,003-005,007)(E)\end{tabular}} & \begin{tabular}[c]{@{}c@{}}T1552 (001,002)(E),\\ T1414(M)\end{tabular} \\ \hline
\end{tabular}%
}
\end{table*}
In order to do a thorough threat analysis for every stage of context-sharing platforms, it is crucial to identify the relevant domains inside the MITRE ATT\&CK framework. In order to identify the pertinent domains, we carried out a comprehensive analysis of all ATT\&CK domains, encompassing enterprise, mobile, and industrial control systems. Our analysis discovered that the fundamental elements of context-sharing platforms encompass all of the MITRE ATT\&CK domains. Therefore, it is essential to thoroughly examine all potential risks related to these areas while doing threat modelling for each stage of context-sharing platforms. An analysis of each tactic in the MITRE ATT\&CK framework is conducted for each stage of the context-sharing platform. Therefore, we define pertinent techniques and sub-techniques linked to each stage of the context-sharing platform. In addition, we create a correlation between the MITRE techniques and sub-techniques and the key elements of the context-sharing platform. Discovering the relevant MITRE techniques and sub-techniques for each element of the context-sharing platform aids in discovering potential mitigation approaches specified in the MITRE ATT\&CK. As a result, this helps to improve the security of the related elements. Due to the large amount of MITRE techniques and sub-techniques that have been identified for threat modelling in context-sharing platforms, we will offer the mapped threats individually. This study focuses on the "credential access" tactic outlined in the MITRE ATT\&CK framework and applies it to the key elements of IoT context-sharing platforms.
\par
Table~\ref{tab:credentialAccess} presents the outcome of applying MITRE ATT\&CK framework (specifically credential access tactic) on IoT context-sharing frameworks. An illustration of how threats are represented in these tables can be seen in the case of the \textit{Multi-Factor Authentication Request Generation (code T1621 (E))} threat. This threat has been identified as a method used in the enterprise domain, specifically under the "credential access" tactics within the ATT\&CK framework. In order to identify the main targets of this approach, we have performed a comprehensive examination of the definitions and descriptions offered by ATT\&CK. Therefore, we have identified the specific components within the context-sharing platform that are relevant for this technique. For example, this technique can be linked to the data consumer element because its main objective is to access the resources of data consumers, such as email. Additionally, it can be associated with the concept of virtualization, as access to VMs or cloud services provided by public and private cloud service providers through Multi-Factor Authentication is a common feature. Based on table~\ref{tab:credentialAccess}, our methodology provides a comprehensive collection of threats, categorised by technique and sub-technique, that are relevant to the initial target of threats in context-sharing platforms. Given that we have reported on a comprehensive list of security risks related to accessing credentials that target different elements inside IoT context-sharing platforms, the corresponding methods to mitigate each identified risk can be found in the MITRE ATT\&CK framework.
\section{conclusion}
This work discussed the importance of developing a systematic and comprehensive threat modelling approach for IoT context-sharing platforms given their key role in IoT deployments and critical infrastructure. We have analysed selected well-known industry-funded projects and identified the main elements of generic IoT context-sharing platforms. We have adopted MITRE ATT\&CK as a well-known and practical framework to conduct a detailed analysis of a specific tactic (i.e., credential access tactic) listing all applicable techniques and sub-techniques. 
Having proved the feasibility of this approach, we plan to conduct a thorough analysis of all tactics outlined in the MITRE ATT\&CK framework and examine potential risks inside IoT context-sharing platforms. A systematic and comprehensive security analysis and threat modelling are necessary to evaluate the security of IoT context-sharing platforms and develop alternate solutions that adhere to the 'secure by design' approach. 
%Additionally, our objective is to provide a threat analysis tool that will assist academics and engineers in evaluating the level of risk associated with their solutions for IoT context-sharing platforms. Additionally, we intend to assess a variety of significant industry projects and research articles using our proposed threat modelling methodology. This will enable us to identify research gaps and emphasise potential future directions.
%
\section*{Acknowledgment}
This work has been supported by the Cyber Security Research Centre Limited (CSCRC) whose activities are partially funded by the Australian Government’s Cooperative Research Centres Programme.

\bibliographystyle{ieeetr}
\bibliography{ref}

\end{document}